\documentclass[pre,twocolumn,groupedaddress,showpacs]{revtex4}
\usepackage{graphicx}
\usepackage{dcolumn}
\usepackage{amsmath}
\usepackage{bm}

\begin{document}
\title{Nematic director slippage: Role of the angular momentum of light}
\author{D. Andrienko$^{1,2,}$, V. Reshetnyak$^{2,3}$, Yu. Reznikov$^{2}$, and T. J.
Sluckin$^{4}$}
\affiliation{$^1$ H. H. Wills Physics Laboratory, University of Bristol, Bristol, BS8 1TL, United Kingdom}
\affiliation{$^2$ Institute of Physics, 46 Prospect Nauki, Kiev, 252022, Ukraine}
\affiliation{$^3$ Physics Faculty, Kiev University, 6 Prospect Glushkova, Kiev, 252022, Ukraine}
\affiliation{$^4$ Faculty of Mathematical Studies, University of Southampton, Southampton, SO17 1BJ, United Kingdom}
\begin{abstract}
We propose a theoretical model of the light-induced director slippage
effect. In this effect the bulk director reorientation contributes to the
surface director reorientation \ It is found that the director and
ellipticity profiles, obtained in the geometric optics approximation, are
dependent on the ellipticity of the incident light wave.\ The director
distribution is spatially modulated in linearly polarized light but grows
monotonically in circularly polarized light. The surface director deviation
has been examined, and comparison made with existing experimental data,
which then permits the magnitude of the orientational nonlinearity
coefficient to be calculated.
\end{abstract}
\maketitle

\section{Introduction}

Light-induced reorientation effects in liquid crystals are important both
for fundamental reasons and because of the possibility of using these
effects in applications \cite{Khoo1993}. In the majority of cases, the
director reorientation occurs in the cell bulk and is imposed by the optical
torque. This torque can originate either from: orientational nonlinearity 
\cite{Zel'dovich1982}, or from photorefractive effects \cite
{Khoo1996,Rudenko1996}, or as a result of photoinduced nonlinearity in
dye-doped LCs \cite
{Janossy1990,Janossy1991,Janossy1992,Janossy1994,Zolot'ko1994,MuensterPRL1997}
.

Light-induced reorientation effects have usually been observed in liquid
crystal cells with {\em strong} anchoring. By strong anchoring we imply that
the director reorientation occurs in the bulk of liquid crystal and is
negligible at the aligning surface. Development of new aligning materials
providing {\em weak} anchoring \cite
{BryanBrown1999,Andrienko1998,Andrienko1999} has allowed the observation of 
{\em surface} director reorientation.

The first observation of such a reorientation, referred hereafter as a {\em 
slippage effect}, was reported by Marusii {\it et al} \cite{Marusii1996} in
a combined liquid crystal cell. One of the cell surfaces with strong planar
anchoring provided homogeneous alignment of the nematic director parallel to
the substrates. The other surface had rather small azimuthal anchoring.
Liquid crystal was doped with azo-dye in order to increase the orientational
nonlinearity coefficient, due to photoinduced trans-cis isomerization of
azo-dye molecules \cite{MuensterPRL1997,JanossyPRE1998}. A linearly
polarized laser beam, propagating through the birefringent mixture, induced
a bulk torque accompanied by director reorientation in the cell bulk and at
the weak anchoring surface. The surface director reorientation was observed
as a change in the polarization state of a probe laser beam propagating from
the side of the strongly anchoring surface.

Recently Francescangeli {\it et al} \cite{Francescangeli1999} found that the
slippage effect competes with a the light-induced anchoring effect
discovered earlier by Voloshchenko {\it et al} \cite{Voloshchenko1994}. At
the same time, was noticed that this effect is sensitive to the polarization
of the laser beam \cite{AndrienkoMCLC1999,AndrienkoUFJ1999}.

The key element of the present paper is to combine the physics of surface
director reorientation with the use of light-induced bulk director
reorientation. The reorientation occurs in a cell in which the director
rotates in an elliptically polarized laser beam, subject to boundary
conditions which constrain the surface director to remain in the plane of
the cell. Light which propagates through a linear birefringent medium in
this geometry changes its polarization state in a periodic way, with
wavelength $\lambda /(n_{e}-n_{o})$. In the course of one wavelength, the
ellipticity of the light changes periodically between -1 and 1, as the light
alternates between linear and circular polarizations. A liquid crystal,
although no longer strictly a linear birefringent medium, in that there is a
feedback on the director from the local fields, nevertheless conserves this
property except for very intense light. As a consequence, the field-induced
torque on the liquid crystal director is also spatially modulated and
periodic. However, if the non-linear feedback increases, the {\em ellipticity
} spatial modulation is accompanied by a {\em director} reorientation. The
reorientation of the director in the cell bulk immediately manifests itself
as a director reorientation on the {\em surface}, if the surface anchoring
is weak.

Since the director torque compensates the deposition of the angular momentum
of light into the liquid crystal, the director reorientation depends
significantly on the polarization state of the incident light beam. As a
result, surface director reorientation is also sensitive to the polarization
of the incident light beam.

The paper is organized as follows. In \S \ref{sec2} we solve the
self-consistent problem of propagation of monochromatic light wave in the
planar liquid crystal cell. We use approximations in the spirit of geometric
optics to obtain the director distribution in the cell. Then in \S \ref{sec3}
we compare existing experimental results to the theoretical predictions and
estimate the photoinduced nonlinearity coefficient. Finally in \S \ref{sec4}
we present some brief conclusions.

\section{The director and ellipticity profiles}

{\label{sec2}} Consider a nematic liquid crystal cell of thickness $L$
confined between the planes $z=0$ and $z=L$ of a cell uniform in the $xy$
plane. Let the director ${\bf n}=(\cos \varphi ,\sin \varphi ,0)$ describe
the average molecular orientation in the cell. We suppose $\varphi =\varphi
(z)$ and only consider the equilibrium orientation of the director, i.e. the
angle $\varphi $ does not depend on time.

Let the elliptically polarized monochromatic light wave, with wave number $
{\bf k}=k{\bf z}$, ellipticity $e_{0}$, and major ellipse axis making an
angle $\alpha $ with the $x$ axis, be normally incident on the cell from the
side of the surface with weak anchoring. Note that, because the light is
propagating in an inhomogeneous anisotropic medium, the Poynting vector and
the light polarization both vary in space throughout the sample.

The light illumination produces a bulk torque acting on the liquid crystal.
The director distribution is determined by the balance between optical,
elastic, and surface torques. To calculate the director configuration ${\bf 
n }(z)$, it is necessary to write down an effective free energy density $f$,
which can then be minimized with respect to the director profile. This is: 
\begin{equation}
f=f_{{\rm el}}+f_{{\rm opt}},  \label{f}
\end{equation}
where $f_{{\rm el}}=\frac{1}{2}K_{22}\left( \partial \varphi /\partial
z\right) ^{2}$ is the nematic liquid crystal elastic energy and $K_{22}$ is
the twist elastic constant.

To obtain an expression for the electromagnetic energy density $f_{{\rm opt}
} $ we suppose the nematic liquid crystal to be a slowly varying uniaxial
medium, so that $\varphi $ varies appreciably only over a length much
greater than the optical wavelength $\lambda $. We also assume that the
birefringence of the medium is small, i.e. $\Delta n=n_{e}-n_{o}<<1$. In
this case the light polarization also varies slowly through the medium, and
we can use the geometrical optics approximation (GOA) to solve Maxwell's
equations for the field in the cell \cite{Ong1983}. We shall also suppose
the medium to be non-absorbing. In this case the beam intensity $I$, defined
as the $z$ component of the average Poynting vector, remains constant.

The approach we shall follow was developed by Santamato {\it et al.} \cite
{Santamato1988,Abbate1991}. We first remind the reader of the elements of
this theory.

The total electromagnetic energy density $f_{{\rm opt}}$ of the light wave
in a non-absorbing, nonmagnetic medium can be rewritten in terms of the
light intensity $I$, ellipticity $e$, and major axis angle $\psi $. In the
limit of low birefringence this takes the form 
\begin{equation}
f_{{\rm opt}}=-\frac{In_{o}}{c}-\frac{I\Delta n}{2c}[1+\left( 1-e^{2}\right)
^{1/2}\cos 2\left( \psi -\varphi \right) ],  \label{f_opt}
\end{equation}
where $I$ is the $z$ component of Poynting vector, or equivalently, the
total intensity of the light wave.

The ellipticity $e$ and the major axis angle $\psi $ can be defined in terms
of the Stokes parameters $\left\{ S_{i}\right\} $, where
\begin{eqnarray}
S_{0} &=&|E_{x}|^{2}+|E_{y}|^{2},  \label{S} \\
S_{1} &=&|E_{x}|^{2}-|E_{y}|^{2},  \nonumber \\
S_{2} &=&2{\rm Re}\left( E_{x}^{\ast }E_{y}\right) ,  \nonumber \\
S_{3} &=&2{\rm Im}\left( E_{x}^{\ast }E_{y}\right) .  \nonumber
\end{eqnarray}

Then the ellipticity, defined by $e=S_{3}/S_{0}$, is the ellipticity of the
polarization ellipse of the light in the cell, and $2\psi =\arctan \left(
S_{2}/S_{1}\right) $ is the angle which the major axis of the polarization
ellipse forms with the $x$ axis.

Equations governing the polarization of light can be derived from the
electromagnetic energy density $f_{{\rm opt}}$ by considering $f_{{\rm opt}}$
as a Hamiltonian function. Then $\psi $ is a generalized coordinate and $%
l_{z}=-(I/\omega )e$ is conjugate momentum, where $l_{z}$ is the average
angular momentum carried by the optical beam along the propagation direction.

These equations are:
\begin{equation}
\frac{\partial \psi }{\partial z}=-\frac{\omega }{2c}\Delta n\frac{e}{\left(
1-e^{2}\right) ^{1/2}}\cos 2\left( \psi -\varphi \right) ,  \label{psi}
\end{equation}
\begin{equation}
\frac{\partial e}{\partial z}=\frac{\omega }{c}\Delta n\left( 1-e^{2}\right)
^{1/2}\sin 2\left( \psi -\varphi \right) .  \label{e}
\end{equation}

The equilibrium orientation of the molecular director is given by the
minimizer of the functional $\int_{0}^{L}f\left\{ \varphi \left( z\right)
\right\} dz$ \ subject to fixed field intensity $I$ and polarization state $%
\left( e,\psi \right) $.

Instead of deriving variational equations it is more useful to use the
conserved quantities corresponding to the Lagrange density $f$. This does
not depend explicitly on the $z$ coordinate. Noether's theorem then implies
that there is a conserved integral analogous to the energy in classical
dynamics. Thus the first integral for this problem can be obtained directly
from the free energy density (\ref{f}):

\begin{equation}
\frac{K_{2}}{2}\left( \frac{\partial \varphi }{\partial z}\right) ^{2}+\frac{
I\Delta n}{2c}(1-e^{2})^{1/2}\cos 2(\psi -\varphi )=E.  \label{energy}
\end{equation}

There is also another conserved integral which can be derived from this free
energy. This is the total angular momentum flux along the propagation
direction, and is the sum of the elastic and optical angular momenta: 
\begin{equation}
K_{2}\frac{\partial \varphi }{\partial z}+\frac{I}{\omega }e=M.
\label{momentum}
\end{equation}
Eliminating the director angle in eq.(\ref{energy}), using eq.(\ref{psi})
and (\ref{momentum}), yields the following equation for the polarization
ellipse: 
\begin{widetext}
\begin{equation}
\left( \frac{\partial e}{\partial z}\right) ^{2}=\left( \frac{\omega \Delta
n }{c}\right) ^{2}\left( 1-e^{2}\right) -\left[ E-\frac{1}{2K_{2}}\left( M- 
\frac{Ie}{\omega }\right) ^{2}\right] ^{2}\left( \frac{2\omega }{I}\right)
^{2}.  \label{e_sq}
\end{equation}
\end{widetext}
This is the fundamental equation governing the distribution of ellipticity
in the liquid crystal cell. Santamato {\it et. al} \cite{Santamato1988},
then used it to discuss the optical Fredericksz transition in a planar cell.
They observed, in addition, that this equation can in general be solved in
terms of elliptic integrals.

We now turn to the application of the equation (\ref{e_sq}) to our geometry.
We note that the physical structure of the general solutions strongly
depends on the constants $M,$ $E$, which in turn depend on boundary
conditions in a rather complicated way. However, for the cell geometry in
which the slippage effect has been observed, considerable simplifications
are possible.

The boundary conditions involve a fixed director orientation at $z=L,$ and a
free but planar boundary at $z=0$: 
\begin{equation}
\left. \varphi \right| _{z=L}=0,\left. \frac{\partial \varphi }{\partial z}
\right| _{z=0}=0.  \label{phi_b}
\end{equation}
Note, that the director can slip over the surface with {\em weak} anchoring
-- no surface torque prevents it from the reorientation. As a result, the
weak anchoring maximizes the response of the director to a bulk field.

The incident light wave is elliptically polarized and initially incident on
the surface with zero anchoring: 
\begin{equation}
\left. e\right| _{z=0}=e_{0}, \left. \psi \right| _{z=0}=\alpha .
\label{e_b}
\end{equation}
In this geometry the constants $E$ and $M$ in eq.(\ref{e_sq}) can be easily
determined from the boundary conditions: 
\begin{eqnarray}
M &=&\left( I/\omega \right) e_{0}, \\
E &=&\frac{I}{c}\left[ n_{o}+\frac{\Delta n}{2}\left[ 1+\left(
1-e_{0}^{2}\right) ^{1/2}\cos 2\left( \alpha -\varphi _{0}\right) \right] 
\right] ,  \nonumber
\end{eqnarray}
where $\varphi _{0}=\varphi (z=0)$. Then, eq.(\ref{e_sq})\ for the
ellipticity and the director (\ref{momentum}) simplify to 
\begin{equation}
\left( \frac{\partial e}{\partial s}\right) ^{2}=1-e^{2}-\left[ \left(
1-e_{0}^{2}\right) ^{1/2}\cos 2\left( \alpha -\varphi _{0}\right) -r\left(
e_{0}-e\right) ^{2}\right] ^{2},  \label{ellipticity}
\end{equation}
\begin{equation}
\frac{\partial \varphi }{\partial s}=r\left( e_{0}-e\right) ,
\label{director}
\end{equation}
where $s=\left( \omega \Delta n/c\right) z=2\pi \Delta n\left( z/\lambda
\right) $. We have also introduced the dimensionless parameter $r$, which is
proportional to the light intensity and defined by: 
\begin{equation}
r=Ic/\left( \Delta nK_{22}\omega ^{2}\right) .  \label{r}
\end{equation}

We now consider the two distinct cases, of linearly and circularly polarized
incident light waves.

\subsection{Linear polarization of the incident light}

Zero ellipticity of the linearly polarized light implies that the light wave
does not carry angular momentum, since the latter is proportional to the
ellipticity.

The initial conditions for the polarization state $e$ read: 
\begin{equation}
e_{0}=0, \left. \psi \right| _{z=0}=\alpha ,
\end{equation}
and the equation for the ellipticity (\ref{ellipticity})\ simplifies to 
\begin{equation}
\left( \frac{\partial e}{\partial s}\right) ^{2}=1-e^{2}-\left[ \cos 2\left(
\alpha -\varphi _{0}\right) -re^{2}\right] ^{2}.  \label{e_sq_lin}
\end{equation}

Substitution $y=re^{2}$ allows (\ref{e_sq_lin}) the standard integral for
the elliptic functions \cite{Gradshtein1994}. After integration, we obtain 
\begin{equation}
e_{{\rm lin}}(s)=e_{0}\mathop{\rm sd}\left( \kappa s\right)  \label{e_lin}
\end{equation}
where $e_{0}=m^{\prime }\sqrt{e_{p}/r},$ $\kappa =\sqrt{e_{p}r}/m$, $%
m^{\prime }=\sqrt{1-m^{2}}$, $m^{2}=e_{p}/\left( e_{p}-e_{n}\right) $, $%
e_{p,n}$ are the positive and negative roots of the equation $1-e^{2}-\left[
\cos 2\left( \alpha -\varphi _{0}\right) -re^{2}\right] ^{2}=0$
correspondingly, $\mathop{\rm sd}(x)=\mathop{\rm sn}(x)/\mathop{\rm dn}(x)$,
where $\mathop{\rm sn}$, $\mathop{\rm cn}$, $\mathop{\rm dn}$ are Jacobi
elliptic functions of index $m$.

The director distribution can be obtained by integrating (\ref{momentum})
and has the form
\begin{equation}
\varphi _{{\rm lin}}(s)=\mathop{\rm arccot}\left( \frac{m}{m^{\prime }} %
\mathop{\rm cn}\left[ \kappa s_{L}\right] \right) - \mathop{\rm arccot}
\left( \frac{m}{m^{\prime }} \mathop{\rm cn} \left[ \kappa s\right] \right)
\label{phi_lin}
\end{equation}
where $s_{L}=\left( \omega \Delta n/c\right) L$.

The estimation of the typical experimental values for $r$ will be done
later. We now find the solution for $r<<1$. 

As it is seen from the angular momentum conservation (\ref{momentum}), to
obtain the director distribution in {\em linear} order in the dimensionless
light intensity $r$ we need to know the ellipticity $e$ in {\em zero} order
in $r$. The latter can be found by solving eq.(\ref{e_sq_lin}) putting $r=0$%
: 
\begin{equation}
e_{{\rm lin}}(s)=\sin \left(2\alpha\right) \sin \left( s\right).
\label{e_lin_sr}
\end{equation}
Then the equation for the director (\ref{momentum}) can be solved using
elementary methods, yielding: 
\begin{equation}
\varphi _{{\rm lin}}(0)=2r\sin \left( 2\alpha \right) \sin ^{2}\left(
s_{L}/2\right) .  \label{phi_lin_sr}
\end{equation}

The ellipticity distribution and the director deviation given in eqs.(\ref
{e_lin_sr},\ref{phi_lin_sr}) might in principle have been obtained by
assuming a priori that the ordinary and extraordinary light waves simply
follow the distribution of the director. Equivalently, one can say that in
the low $r$ limit the light beam propagates through the cell in the
adiabatic or Mauguin regime \cite{deGennes1993}. 
In the adiabatic approach one can write the expression for the electric
field components explicitly since the ordinary and extraordinary light waves
follow the distribution of the director. Then this field is used as an
external field which reorients the director. However, such an assumption is
not theoretically satisfactory, and naive use of this approximation as an
initial hypothesis can lead to internally inconsistent approximations \cite
{Ong1983a}.

Let us now summarize the important properties of the light propagation and
the director distribution in the planar cell. Eq.(\ref{e_lin_sr}) yields
that the polarization ellipse $e(z)$ is a periodic function with spatial
period $l=\lambda /\Delta n.$ In more general case (\ref{e_lin}) the spatial
period depends on the light intensity, $l=4K\left( m\right) \left( \kappa
\omega \Delta n/c\right) ^{-1}$, where $K\left( m\right) $ is the complete
elliptic integral of the first kind.

\begin{figure}[tbp]
\includegraphics[width=8cm, angle = 0]{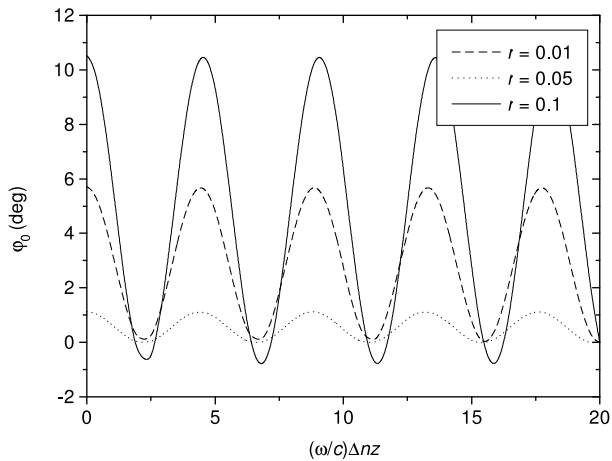}
\caption[Fig 1]{Director distribution in the cell for different values of
the dimensionless light intensity $r$. Linear polarization of the incident
light. The director angle varies periodically through the cell thickness
following the change in the light ellipticity.}
\label{fig:1}
\end{figure}

The liquid crystal director $\varphi (z)$ follows the distribution of the
ellipticity with the same spatial period, and is shown in Fig.~\ref{fig:1}.
It is seen that, in spite of the modulation, the average deviation of the
director tends to be along the polarization of the incident light and
increases with the light intensity.

We now make some comments on the director deviation $\varphi _{0}$ at $z=0$.
First, eq. (\ref{phi_lin_sr}) shows that the surface director is sensitive
to the sign of the birefringence $\Delta n$. If $\Delta n>0$, the director
tends to be parallel to the polarization direction. It reorients
perpendicular to the polarization direction if $\Delta n<0$.

\begin{figure}[tbp]
\includegraphics[width=6.0cm, angle = -90]{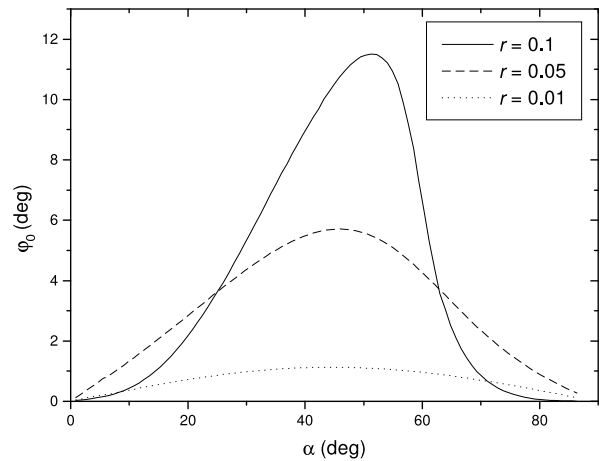}
\caption[Fig 2]{Director deviation on the isotropic surface as a function of
the angle between rubbing direction and polarization of the exciting light.
The director response has a maximum at $\protect\alpha = \protect\pi / 4$
and is absent at $\protect\alpha = 0, \protect\pi / 2$ for small light
intensities.}
\label{fig:2}
\end{figure}

Second, in the limit of small intensities, $r<<0$, the liquid crystal
response is maximal at $\alpha =\pi /4$, and does not occur for $\alpha
=0,\pi /2$ since we are below the threshold of Freederiksz transition.
However, in the general case, the value of $\alpha $ which provides maximal
director response depends on the light intensity. For bigger values of the
parameter $r$ the maximal response of the director occurs at an angle $%
\alpha $ larger than $\pi /4$ (Fig. \ref{fig:2}).

The most important comment is that the director deviation depends strongly
on the cell thickness $L$ and refractive index difference $\Delta n$.
Indeed, $\omega \Delta nL/\left( 2c\right) =\pi \Delta nL/\lambda ,$ and for
typical experimental conditions $L/\lambda \approx 100$. Thus, a change in
the refractive index $\Delta n\sim 10^{-2}$ leads to a considerable change
in the director deviation $\varphi _{0}$. In practice this means that
changes in the refractive index due to the laser-induced heating of the
liquid crystal \cite{Khoo1988} or molecular phototransformations \cite
{Pinkevich1992} influence the amplitude of the surface director deviation.
This also means that the cell thickness should be tuned rather precisely in
order for the surface director orientation to be observed. The optimal
condition is that $\Delta nL/\lambda =N/2$, where $N$ is an integer.

\subsection{Circular polarization of the incident light}

Contrary to the case of linear polarization, circularly polarized light wave
carries angular momentum along the propagation direction. The initial
conditions for the polarization state $e$ are now: 
\begin{equation}
e_{z=0}=1.
\end{equation}
The boundary conditions for the director are the same as for the linearly
polarized light (\ref{phi_b}).

The equation for the ellipticity then simplifies to 
\begin{equation}
\left( \frac{\partial e}{\partial s}\right) ^{2}=1-e^{2}-r^{2}\left[ 1-e %
\right] ^{4}.
\end{equation}
Substitution $y=1-e$ allows integration of this equation in terms of
elliptical integrals \cite{Gradshtein1994}. After integration we get the
distribution of the ellipticity in the cell:

\begin{equation}
e_{{\rm cir}}(s)=1-e_{r}{{\frac{q\left[ 1-\mathop{\rm cn}\left( \kappa
s\right) \right] }{p+q+\left( p-q\right) \left[ 1-\mathop{\rm cn}\left(
\kappa s\right) \right] }}}  \label{e_cir}
\end{equation}
where $e_{r}$ is the only real root of the equation $2-e-r^{2}e^{3}=0,$ $%
p^{2}=r^{-2}+3e_{r}^{2},$ $q^{2}=r^{-2}+e_{r}^{2},$ $\kappa =r\sqrt{pq}$.
The polarization ellipse $e_{{\rm cir}}(s)$ has the same dependence on the
spatial coordinate as we have in the case of linear polarization (\ref{e_lin}
), i.e. is a periodic function with the same spatial period.

Now we turn to the director distribution. Integrating Eq. (\ref{momentum})
we obtain 
\begin{equation}
\varphi \left( s\right) =\varphi _{0}+\frac{e_{r}}{p-q}\left[ 2\arctan
\left( \sqrt{\frac{q}{p}}\tan \left( \frac{\kappa s}{2}\right) \right) - 
\sqrt{\frac{q}{p}}\kappa s\right] .  \label{phi_cir}
\end{equation}
To perform the integration we used an approximation $\mathop{\rm cn}
(x)\simeq \cos (x)$, which is valid for small enough elliptic function
indices or alternatively small enough values of $r$.

The expressions (\ref{e_cir},\ref{phi_cir}) in leading order in $r$ are:
\begin{equation}
e_{{\rm cir}}\left( s\right) =\cos \left( s\right),  \label{e_cir_sr}
\end{equation}
\begin{equation}
\varphi _{{\rm cir}}(0)=r\left[ s_{L}-\sin \left( s_{L}\right) \right].
\label{phi_cir_sr}
\end{equation}

Let us now analyze the ellipticity and the director profile in the cell. Eq.
(\ref{e_cir_sr}) yields that polarization ellipse is exactly the same as for
the linearly polarized light. The only change is that, in order to match the
condition $e_{{\rm cir}}\left( 0\right) =1$ instead of $e_{{\rm lin}}\left(
0\right) =0$, we have to have cosine dependence on $s$ (cf. exp.(\ref
{e_lin_sr})).

\begin{figure}[tbp]
\includegraphics[width=6cm, angle = -90]{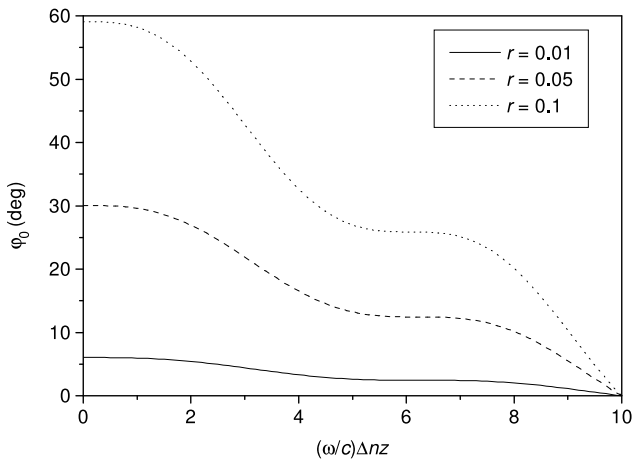}
\caption[Fig 3]{ Director distribution in the cell for different values of
the dimensionless light intensity $r$. Circular polarization of the incident
light. The director angle increases momotonically through the cell
thickness. }
\label{fig:3}
\end{figure}

However, the liquid crystal director is no longer a periodic function of the
cell thickness. Even more, now it is {\em monotonically growing }function of 
$s$ (Fig. \ref{fig:3}). This fact can be easily proved if one considers the
angular momentum conservation (\ref{momentum}). For the {\em linear}
polarization it reads: $\partial \varphi /\partial s=-re$. Since $e\left(
s\right) $ is a periodic function of $s$, $\varphi $ is also a periodic
function of $s$ with the same spatial period. However, for {\em circular}
polarization we have $\partial \varphi /\partial s=r(1-e)\geq 0$. Therefore, 
$\varphi $ grows monotonically with $s$. In fact, the director distribution
is a superposition of monotonically growing and oscillating functions, since
the ellipticity of the polarization ellipse\ is a periodic function of $s$.

It is also seen from eq. (\ref{phi_cir_sr})\ that the director deviation $%
\varphi _{0}$ at $z=0$ consists of two parts. The first part, $r\sin \left(
s_{L}\right) $, originates from the ellipticity modulation. The other part, $%
-rs_{L}$, is due to the constant injection of light angular momentum into
the cell bulk. For typical experimental conditions this contribution
dominates, since $s_{L}=2\pi \Delta nL/\lambda >>1>\sin \left( s_{L}\right) $
. For the same reasons $\varphi _{0}^{{\rm cir}}\simeq 2\pi \Delta
nL/\lambda \varphi _{0}^{{\rm lin}}$. Therefore, the surface director
deviation in circularly polarized light is $2\pi \Delta nL/\lambda $ times
bigger than in the linearly polarized light.

We can make even more interesting remarks. 

First, the resulting twist angle $\varphi _{0}$ is not sensitive to the
variation of $\Delta n$ for typical experimental conditions ($r << 1$) and
liquid crystal cell parameters ($s_L >> 1$). This is very different from the
situation with linearly polarized light wave. Second, $\varphi _{0}$ is
proportional to the cell thickness $L$, i.e., in thick enough cells, its
value can exceed $\pi $ and one can get supertwisted distribution of the
director or even distribution that realizes in chiral nematics.

\subsection{Elliptically polarized light}

Now we try to solve the eqs. (\ref{ellipticity},\ref{director}) for the case
of elliptically polarized light. Solving these equations for arbitrary value
of $r$ is a complicated task. However, as one can see from the angular
momentum conservation (\ref{momentum}), to obtain the director distribution
in {\em linear} order in the dimensionless light intensity $r$ we need to
know the ellipticity $e\left( s\right) $ in {\em zero} order in $r$. Solving
eq.(\ref{ellipticity}) in this limit we obtain that $e\left( s\right) $ is a
superposition of the solutions $e_{{\rm lin}}\left( s\right) $ and $e_{{\rm %
cir}}\left( s\right) $: 
\begin{widetext}
\begin{equation}
e\left( s\right) =\sqrt{1-e_{0}^{2}}\sin \left( 2\alpha \right) \sin
s+e_{0}\cos s=\sqrt{1-e_{0}^{2}}e_{{\rm lin}}\left( s\right) +e_{0}e_{{\rm %
cir}}\left( s\right)
\end{equation}
Correspondingly, in the leading order in $r$, the director deviation at $z=0$
is also a superposition of $\varphi _{0}^{{\rm lin}}$ and $\varphi _{0}^{%
{\rm cir}}$ and has the form: 
\begin{equation}
\varphi _{0}=r\left[ \sqrt{1-e_{0}^{2}}\sin \left( 2\alpha \right) \left(
1-\cos s_{L}\right) +e_{0}\left( \sin s_{L}-s_{L}\right) \right] =\sqrt{
1-e_{0}^{2}}\varphi _{0}^{{\rm lin}}+e_{0}\varphi _{0}^{{\rm cir}}
\label{phi_ell}
\end{equation}
\end{widetext}
It is seen from Eq. (\ref{phi_ell}) that, for circularly polarized light
wave, the director at the surface rotates according to the sign of the
ellipticity: clockwise polarization leads to the clockwise reorientation of
the director and vice versa.

\section{Discussion}

{\label{sec3}} We now turn to the estimation of the parameter $r$ and the
director deviation angles one can expect for typical experimental conditions.

Quantitatively, in typical experiments, we have light intensities $%
I<10W/cm^{2}$, and liquid crystal constants $\Delta n=0.1$, $K_{22}=3\cdot
10^{-7}dyn$. With these fundamental parameter values, the dimensionless
intensity $r<0.001<<1$, which is fairly small. The maximal deviation of the
director which can be observed under these circumstances is far less than $
0.1^{0}$ for linearly polarized light and is less than $1^{0}$ for
circularly polarized light. Such reorientations can hardly be detected in
real experiments.

However, the situation is different for the{\em \ dye-doped }liquid
crystals. All previous experimental results \cite{Janossy1990,Janossy1991}
and theoretical predictions \cite{Janossy1994,Marrucci1997}\ are consistent
with the following expression for the dye contribution to the free energy
density: 
\begin{equation}
f_{{\rm dye}}=\eta f_{{\rm opt}},  \label{f_dye}
\end{equation}
where $f_{{\rm opt}}$ is the total electromagnetic energy density of the
incident light wave (\ref{f_opt}) and the parameter $\eta $ characterizes
the efficiency of the dye-induced director reorientation. The parameter $
\eta $ is proportional to the dye concentration, but also depends on the
molecular structures of both dye and liquid crystal.

Eq.(\ref{f_dye}) implies that the effective torque imposed by the light wave
increases proportionally to $\eta $. In typical experiments, we obtain a dye
assisted nonlinearity coefficient $\left| \eta \right| \simeq 200$ \cite
{Janossy1992}. This gives $r\simeq 0.1$ and typical director deviations
about $10^{0}$, which can be measured experimentally without any
difficulties.

Our theoretical framework also permits the evaluation of the photo-induced
orientational nonlinearity coefficient $\eta $ from the existing
experimental data. $\eta $ is the crucial phenomenological parameter which
governs the interaction between light and the liquid crystal/azo-dye
mixture. This quantity is an input parameter in our theory, and must either
be measured or calculated using a microscopic theory \cite
{Janossy1994,Marrucci1997}.

Trying to estimate $\eta $ we are fully aware that the cell is absorbing,
since the effect of the photoinduced nonlinearity is due to the absorption
of light by the dye molecules. For the absorbing media, we are no longer 
able to apply the Lagrange approach, which involves energy and angular 
momentum conservation. Therefore, estimating coefficient $\eta$ we are 
neglecting effects related to the light absorption.

For the linearly polarized incident light, the value of $\eta $ can be
determined applying eq.(\ref{phi_lin_sr}), which connects $\varphi _{0}$ to
the dimensionless intensity $r$, to the experimental results presented in 
\cite{Marusii1996}. 

Typical experiments were performed in the cell with one isotropic surface
and one surface consisting of a substrate covered with a rubbed polyimide
film. The cell was filled with the nematic liquid crystal
4-n-pentyl-4-cyanobiphenyl (5CB) doped with the azo dye methyl red (MR) at
weight concentration $0.1$. The mixture was exposed from the side of the
isotropic surface by a linearly polarized beam of a cw Ar$^{+}$ laser,
irradiation wavelength of which $\lambda =488{\rm nm}$ is near the maximum 
of the absorption band of MR dissolved in 5CB. Typical intensity of the
 Ar$^{+}$ laser was in the range $0.1-1{\rm W}/{\rm cm}^{2}$. Its
polarization vector was set at $45^{o}$ with respect to the initial director 
orientation. The cell response was observed with a probe He-Ne laser beam with 
wavelength $\lambda =638{\rm nm}$ which is in the region of MR transparency. 

\begin{figure}[tbp]
\includegraphics[width=6cm, angle = -90]{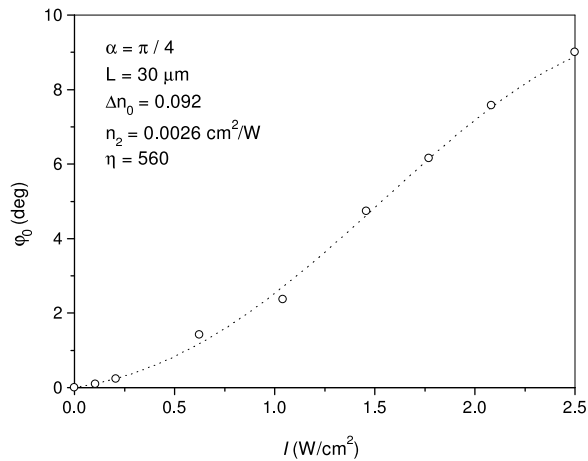}
\caption[Fig 4]{Experimental data and the best fit of the director angle $%
\protect\varphi _{0}$ as a function of the exciting laser beam intensity ($%
\protect\alpha =\protect\pi /4$). Fitting parameters: photoinduced
nonlinearity coefficient $\protect\eta $ and refractive index of the mixture 
$\Delta n_{0}$.}
\label{fig:4}
\end{figure}

Irradiation with a beam of a Ar$^{+}$ laser led to the director
reorientation in the liquid crystal bulk and its slippage on the isotropic 
surface. The latter was  detected as a rotation of the polarization of the probe beam. 
The experimental dependence of the rotation angle on the intensity
of the incident light is presented in Fig.\ref{fig:4}.

Eq. ({\ref{phi_lin_sr}}) predicts that the surface director reorientation
angle $\varphi _{0}$ will be a function of the laser intensity $I$ and
refractive index anisotropy $\Delta n=n_{e}-n_{o}$. The birefringence of the
dye-liquid crystal mixture can in principle also depend on the exciting
light intensity indirectly both as a result of laser-induced heating of the
liquid crystal \cite{Khoo1988}, and because molecular phototransformation
can change the dipole moment of the phototransformed molecules \cite
{Pinkevich1992}. The intensity $I$ thus has a direct and an indirect effect
on the reorientation of the director. In order to interpret the results
correctly it is necessary to correct for the indirect effects. To do this we
studied the dependence of the dye-liquid crystal birefringence $\Delta n$ on
the exciting light intensity independently using a light-induced
birefringence technique \cite {Pinkevich1992}. We found that in our
case: 
\begin{equation}
\Delta n=\Delta n_{0}+n_{2}I,  \label{deltaneq}
\end{equation}
where $n_{2}\simeq 0.0026\left( {\rm W/cm}^{2}\right) ^{-1}$.

Then, fitting of the data (Fig. \ref{fig:4}) yields $\eta =-560$. The
photo-induced orientational nonlinearity is considerably enhanced even with
respect to the giant orientational non-linearity of pure liquid crystals 
\cite{Zel'dovich1982} and is consistent with what might be expected as a
result of measurements in related systems \cite{Zolot'ko1994,JanossyPRE1998}
using the $z$-scan technique.

\begin{figure}[tbp, ]
\includegraphics[width=6cm, angle=-90]{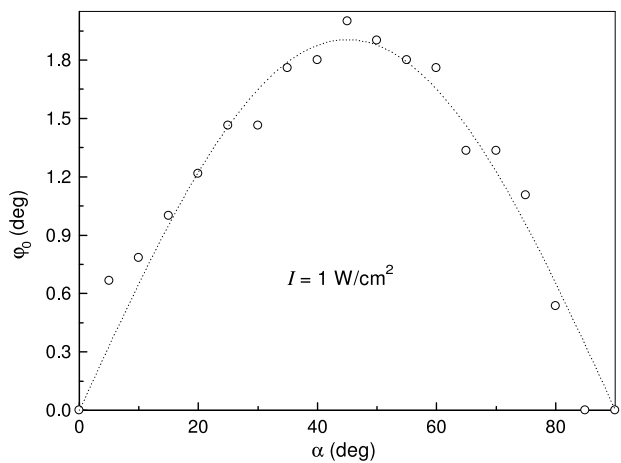}
\caption[Fig 5]{Director deviation on the isotropic surface as a function of
the angle between rubbing direction and polarization of the exciting light.
The director response has a maximum at $\protect\alpha =\protect\pi / 4$ and
is absent at $\protect\alpha =0, \protect\pi / 2$.}
\label{fig:5}
\end{figure}

The theory developed in eq.({\ref{phi_lin_sr}) also requires that the
director response to the laser illumination light intensity be linear. In
order to check whether this is the case, we measured the dependence $\varphi
_{0}\left( \alpha \right) $. The results of measurements for $I = 1 {\rm W/cm%
}^{2}$ are presented in Fig.\ref{fig:5}. We find, in agreement with eq. (\ref
{phi_lin_sr}), that the liquid crystal response is maximal at $\alpha =$}${%
\pi /4}${\ and absent at $\alpha =0,$}${\pi /2}$.

As far as we are aware, there have been no direct observations of the 
slippage effect for circularly or elliptically polarized light. The effect 
has probably been implicitly observed in azo-dye doped liquid crystal cells 
\cite{AndrienkoMCLC1999,AndrienkoUFJ1999}.  In these experimetns it 
was  found  that  when a cell is irradiated by circularly
polarized light, an easy orientation axis is induced on an initially isotropic 
aligning surface. The direction of this axis correlated with the direction 
of the polarization rotation.
The appearance of the easy axis can be explained as follows. First there is 
a light-induced surface director slippage. The director then freezes as a 
result of light-induced {\em adsorption} of the dye molecules on the aligning 
surface.  By contrast, this reorientation cannot be explained if we use a 
model which depends only on the selective {\em adsorption} of the 
dye molecules on the aligning surface.

\section{Conclusions}

\label{sec4} In this paper we have tried to interpret existing experimental
facts on the surface director reorientation by assuming the dye-induced
torque to be proportional to the optical torque, and treating the nonlinear
interaction between the light wave and the mixture in the geometrical optics
approximation.

We now turn to a brief discussion of the implications of this work. We
believe that the experimental technique we introduce in this paper can be
used in further studies of the liquid crystal interface. Relatively
straightforward extensions of the technique will, for example, permit
measurements of fundamental surface parameters such as in-plane surface
viscosities and weak anchoring coefficients.

There are also possible applications of the method involving the writing of
the dynamic holographic polarization gratings. Furthermore, the present
investigation, together with our results published earlier \cite
{Francescangeli1999,Voloshchenko1994,Simoni1997,Slussarenko1997} have shown
that it is possible to control the light-induced alignment memory effect by
changing the concentration of azo-dye in the liquid crystal. There may also
be scope for systematic investigations of surface memory effects, the basic
mechanisms of which are still little understood.

\acknowledgments

D. A. acknowledges support by EPSRC and support through the Grant No.
ORS/99007015 of the Overseas Research Students Awards scheme.

\end{document}